\documentstyle[preprint,aps]{revtex}
\begin{document}
\preprint{\today}
\title{  Inverted Pairing and Excitation Induced by Quadrupole-Quadrupole
Interaction}
\author{Hsi-Tseng Chen and Cheng-Li Wu}
\address {Department of Physics, Chung-Yuan Christian University\\
Chungli, Taiwan, ROC}
\author { Da Hsuan Feng}
\address {Department of Physics and Atmospheric Science, Drexel University\\
Philadelphia, PA 19104-19104, USA }
\author {Akito Arima}
\address {RIKEN\\
Harosawa 2-1, Wako, Saitama 351-01, JAPAN}

\maketitle
\begin{abstract}
In this paper, eit is shown for a single-$j$ shell that the intrinsic
excitations induced by the quadrupole-quadrupole interaction can be viewed as
that due to the inverted
surface-delta interaction whose multipole-pairing component with the largest
spin is the dominant term and is attractive. This allow us to trace the origin
of the recently introduced optimum pair description of rotation  and leads
naturally to analytical expressions for the rotational energies for the various
bands in a single $j$-shell with the
quadrupole-quadrupole interaction \end{abstract}

\newcommand{\jja}{j(j+1)(2j-1)(2j+1)(2j+3)}
\newcommand{\jjx}{j(j+1)(2j+1)(2j+3)}
\newcommand{\jj}{\Lambda}
\newcommand{\jjb}{j(j+1)}
\newcommand{\jjc}{\Lambda(\Lambda+1)}
\newcommand{\jjd}{\Lambda^2(\Lambda+1)^2}
\newcommand{\jju}{(2    L'_{1}   +1)(2  L'_{2}  +1)X^2(jj
                 L_{1}  ;jj  L_{2} ;  L'_{1}L'_{2} L)}

\clearpage
\hspace{0.5cm}    One of the central aims of nuclear structure physics is to
understand how strong collectivity is manifested in the spherical shell model.
Since straightforward application of the shell model is not possible for
strongly deformed systems, a great deal of effort in the past forty years was
devoted to the development of suitable truncation schemes to confront this
problem ${}^1$,${}^2$.

It is well known now that under a strong monopole pairing interaction plus
a weak quadrupole-quadrupole (QQ) interaction, monopole and quadrupole
pairs appear to be a good truncation scheme.  However, when the QQ interaction
becomes dominant, it appears that higher multipoles, especially the
hexadecapole pairs, must be added ${}^3$.

On the other hand, it has been recently been shown that the eigenstates of the
pure QQ interaction in a single-$j$ shell can be well described in terms
of a collection of optimum pairs, using a novel approach which was
proposed to describe nuclear states within a spherical shell model with
an optimum pair basis${}^4$.  This paper aims to trace the origin
of such a description of rotation in a single-$j$ shell with the
strong QQ interaction.  The physics behind
this approach can be easily understood as follows.  The
major part of the QQ interaction in the single $j$ shell can be
expressed as a constant plus an $L^2$ term.  The remaining parts can be
simulated by an inverted Surface-Delta-Interaction (SDI), which will
favor higher spin pairs.  The constant plus the $L^2$ terms can be
exactly diagonalized.  This will give the rotational spectrum. The
remaining part (i.e. the inverted SDI) can be approximately
diagonalized by using first the standard many-body Tamm-Dancoff approximation,
followed by the optimum pair approach.  A by-product of this study
is the derivation of analytical expressions for the rotational
energies for various bands, for a system of any number of pairs moving
in a single$j$ shell, although the number dependence is not understood
at this point.  In hindsight, this approach is possible because the
intrinsic excitations induced by the QQ-interaction comes essentially
from the inverted SDI.

We shall illustrate our method via a simple solvable model: like-particles
moving in a single-$j$ shell. This example is chosen because the QQ-interaction
can simulate the rotational excitation structure${}^2$.  For a system of $n/2$
or N pairs of nucleons in single-$j$ shell, the most general Hamiltonian is
\begin{equation}
{\bf H}=\epsilon_j\sqrt{2j+1} ({\bf a}_{j}^{\dag}\times{\bf \tilde
a}_{j})^0
+ \sum_{\Lambda}\sqrt{2\Lambda+1}C_{\Lambda}
({\bf A}_{\Lambda}^{\dag}\times{\bf\tilde  A}_{\Lambda})^0
\end{equation}
with
\begin{equation}
{\bf A}_{\Lambda}^{\dag}=\frac{1}{\sqrt{2}}
({\bf a}_{j}^{\dag}\times{\bf a}_{j}^{\dag})^\Lambda
\end{equation}
and
\begin{equation}
C_{\Lambda}=<j j \Lambda\parallel{\bf V}\parallel j j \Lambda>
\end{equation}
where ${\bf  a_{j}^{\dag}}$ (${\bf \tilde a_{j}}$) are the creation
(annihilation) operators with energy ${\epsilon_j}$.

Using the equation-of-motion technique${}^5$, we shall employ the many-pair
creation operator as the excitation operator and the Tamm-Dancoff vacuum as the
"true" vacuum. To  portray  a collection of multipole pair excitations, the
excitation operator is expressed as a simple tensor product:
\begin{equation}
{\bf  \Omega^{\dag}}={\bf  A}_1^{\dag}  {\bf A}_2^{\dag}....{\bf
A}_N^{\dag}
\end{equation}
where   $   {\bf   A}_{\mit   i}^{\dag}$  stands   for   $  {\bf
A}^{\dag}_{L_{\mit i}}$ , the tensor of rank $L_{\mit i}$.

Following the standard process of linearization,the corresponding
equation-of-motion becomes
\begin{equation}
[{\bf H},{\bf \Omega^{\dag}}]=E_{T}{\bf \Omega^{\dag}}
\end{equation}
and the TDA energies are
\begin{equation}
E_{T}=\sum_{\Lambda}N_{\Lambda}C_{\Lambda}
\end{equation}
Here $N_{\Lambda}$ is the number of such pairs with angular momentum $\Lambda$.

It is known that the TDA is most valid when $j$ is large and for interactions
such as surface delta where the strengths of the multipole pairing components
decrease monotonically.  It is unsuitable for the identity interaction and the
QQ force. For example, one can easily show that the TDA ground state energy for
four particles in $j$ = 21/2 orbit is only a third (a fifth) of the exact value
 for
the identity  (QQ) interaction. However, as will be shown later, the inadequacy
of the TDA for these interactions can be salvaged by the introduction of an
auxliary Hamiltonian to the QQ force which will render it as an "inverted" SDI
interaction, one in which the strengths of the new multipole pairing components
increase rather than decrase monotoniclly.  It will restore the validity of the
TDA. To this end, we shall introduce the following auxiliary Hamiltonian as
\begin{equation}
H'=H_{QQ}-\alpha \sum_{i<j}I_{ij}-\gamma {\bf L}^{2}
\end{equation}
where ${\bf L}$ is the angular momentum operator and $I_{\mit ij}$  stands for
the identity interaction and   ${\bf  L}$ and $I_{\mit ij}$ are so chosen for
their simplicity in representing an exact symmetry of the system. In fact, the
auxiliary Hamiltonian  of Eq.(7) has the same form as the original one given by
Eq.(1) if the interaction matrix element, or multipole-pairing strength, are
redefined as
\begin{equation}
C'_{\Lambda}=C_{\Lambda}  - \alpha - \gamma [\Lambda(\Lambda+1)- 2j(j+1)]
\end{equation}
where the parameters $\alpha$ and $\gamma$ in (8) are determined by expressing
this new Hamiltonian as an inverted SDI with the following choice
\begin{eqnarray}
\alpha & = &
\frac{10-j(j+1)}{7}{\bf C}_{2}+\frac{j(j+1)-3}{7}{\bf C}_{4}
\nonumber\\
\gamma & = & \frac{1}{14}({\bf C}_{4}-{\bf C}_{2})
\end{eqnarray}
Explicitly, they are
\begin{equation}
\alpha = \frac{20(j-2)(j+3)(2j-3)(2j+5)}{ \jja }
\end{equation}
\begin{equation}
\gamma = \frac{15[4j(j+1)-27]}{ \jja }
\end{equation}

Using the above parameters, one can show that the multipole pairing strength
$C'_{\Lambda}$ are
\begin{equation}
C'_{\Lambda}=-\frac{15( \jj -2)( \jj +3)( \jj -4)( \jj +5)}{ \jja}
\end{equation}
The invertness can also explicitly be seen numerically. For j=21/2, the values
of multipole-pairing strengths are \[
[-0.001,0,0,-0.0093,-0.04,-0.11,-0.24,-0.46,-0.79,-1.27,-1.95] \]
for $\Lambda=0$,$\Lambda=2$,$\Lambda=4$,.....,$\Lambda=20$.

Applying TDA to the inverted SDI, we can obtain the following energy
expression:
\begin {equation}
E=\sum_{\Lambda}N_{\Lambda}C'_{\Lambda}+\frac{n(n-1)}{2}\alpha
+\gamma[L(L+1)-nj(j+1)]
\end{equation}

Here the TDA energies,$\sum_{\Lambda}N_{\Lambda}C'_{\Lambda}$ are degenerate
for all $L$ values and thus serve as the intrinsic states from which the
various rotational bands are built. The intrinsic excitations are the pairs
with the energy $C'_{\Lambda}$.  The expression  (13) with the numerical values
of (12) clearly indicate that the pairs with the lowest energy does indeed have
the highest spin, and hence the system's ground state must be due to the
contribution  of N pairs of nucleons, each coupled to the highest spin ($  \jj
=  2j-1$). This is of course precisely the behavior demonstrated in a pair
shell model calculation with the  optimal pair approximation${}^3$. For
the two pair case, the ground band is labeled as  $(2j-1;2j-1)$ since it is
formed by two $2j-1$ pairs and the spin sequence of the band is
$0,2,4,6,....,$
while the first excited band is a (2j-3;2j-1)-band (one $2j-3$ pair and one
$2j-1$ pair) with the $\gamma$-band spin sequence of $2,3,4,5,....$. Finally,
the  $\beta$-band  comes from coupling two $2j-3$-pairs.

In Table I, results for a system of four particles in a $j$ = 21/2 orbit are
presented. In particular, the above approximation is compared with the exact
shell model calculations.  In this approximation, the  values of $C'_{\Lambda}$
are given by (12), referred to as Approx.\ I. This distinguishes it from the
other choices of $C'_{\Lambda}$ mentioned below. In view of the simplicity of
the model, the agreement is quite good.

It is interesting to note that the above choice of $C'_{\Lambda}$ is close
to that of fixing $\alpha$ and $\gamma$ in (8) which
result in cancelling the
$a+b\Lambda(\Lambda+1)$ terms in the $C_{\Lambda}$ for QQ interaction. This
latter is given as $10W(jjjj;2\Lambda)$ and can be expanded in
$\Lambda(\Lambda+1)$:
\begin{equation}
C_{\Lambda}=a+b\Lambda(\Lambda+1)+c\Lambda^2(\Lambda+1)^2
\end{equation}
where
\[a=-\frac{10}{2j+1}\]
\[b=\frac{15(4j(j+1)-1)}{ \jja }\] \[c=-\frac{15}{ \jja }\]

The $\alpha$ and $\gamma$ thus determined are
\begin{equation}
\alpha=\frac{80j(j+1)}{(2j-1)(2j+1)(2j+3)}
\end{equation}
\begin{equation}
\gamma=\frac{15(4j(j+1)-1)}{ \jja }
\end{equation}
and the new multipole pairing strengths are
\begin{equation}
C'_{\Lambda}=-\frac{15 \jj ^2( \jj +1)^2}{ \jja }
\end{equation}
which are numerically close to that given by (12) (See Table II).

The above results imply that one can decompose the multipole pairing strengths
in the QQ-interaction into the collective part $a+b\Lambda(\Lambda+1)$ and the
intrinsic part $c\Lambda^2(\Lambda+1)^2$. The latter contributes to the
intrinsic energies of TDA.  The approiximation led by this choice of
$C'_{\Lambda}$ is given as Approx.II in Table I.

Besides the above choices of $\alpha$ and $\gamma$ which led to the inverted
SDI-like interactions , an additional one is added in which the parameters are
fixed so that $C'_{\Lambda}$ is as close to an inverted SDI as possible.  A
strict inverted SDI is defined as
\begin{equation}
C_{ \jj }^{IV} = \frac{2}{2j+1}C_{2j-1- \jj }^{SDI}
\end{equation}
and the  $\chi^2$-fit treatment would lead to the following expressions for
$\alpha$ and $\gamma$:
\begin{equation}
\gamma  = \frac{<(C_{ \jj }-C_{ \jj }^{IV}) \jjc >-<(C_{ \jj }-C_{
\jj }^{IV})>< \jjc >}{< \jjd >-< \jjc >^2}
\end{equation}
\begin{equation}
\alpha = <(C_{ \jj }-C_{ \jj }^{IV})>-\gamma[< \jjc >-2j(j+1)]
\end{equation}
where $<....>$ represents the average taken as
\[\frac{\sum_{  \jj =0}^{\frac{2j-1}{2}}(2 \jj +1)(....)}{\sum(2
\jj +1)}\]
When these are applied to the QQ force, the resulting analytical expressions
are complicated and not physically transparant.  If one choose the following
simple expressions (which will fit (19) and (20) well),
\begin{equation}
\alpha = \frac{80(j-3)(j+4)(j-5)(j+6)}{ \jja }
\end{equation}
\begin{equation}
\gamma = \frac{60(j-5)(j+6)}{ \jja }
\end{equation}
one can obtain the results as presented in Tables I and II (denoted as Inverted
SDI).

The comparisons we presented in Table  I with the various choices of
$C'_{\Lambda}$ suggest that the TDA with the inverted-SDI  is an excellent
caricature of the exact shell  model calculations for the single-j case.
Consequently, the  intrinsic  excitations induced by the  QQ-interaction are
described  according to our simple model (13) as elementary with energies
$C'_{\Lambda}$, from which all the rotational bands are built upon, which are
the multipole-pairing strengths in an inverted SDI.  We therefore conclude
from (8)  that  the  "intrinsic" part of the QQ force is equivalent to an
inverted SDI.  With this view point, one can understand the origin of the
optimum  pair description of rotation. In physical terms, the short  range
delta force will favor the S-pairs,  the  QQ force  favors high spins pairs
whose field is generated by an inverted SDI.

The present model given by Eq. (13) has two shortcomings.  The first is the
moment  of  inertias  are identical for all bands and the second is that the
parameter $\alpha$ is independent of the elementary excitations .  Using a more
rigorous  treatment,(shown next) for the TDA for arbitrary interactions, the
above results can be improved.  In this treatment, which is akin to the
generalized pair field method (GPFM)${}^4$, the results for the QQ interaction
are nearly exact and more importantly, the moment of inertia can vary from
band
to band.

We thus begin by considering the commutator of the auxiliary Hamiltonian
\begin{equation}
{\bf H'}={\bf H}-\omega \sum_{i<j}I_{ij}
\end{equation}
and ${\bf \Omega^{\dag}}$ in (4). {\bf H'} can formally be expanded as
\begin{equation}
[{\bf H'},\Omega^{\dag}]=E'_{T}{\bf \Omega^{\dag}}\nonumber\\
+\sum_{1'2'...N'}\Phi(12...N;1'2'...N';\omega)
({\bf A}_{1'}^{\dag}{\bf A}_{2'}^{\dag}....{\bf A}_{N'}^{\dag})^L
\end{equation}
where $\omega$ is chosen by a $\chi^2$-fit with only the first term surviving.
Again, the TDA is made valid by the proper choice of the new Hamiltonian!

For a system of four particles moving in a single j-orbit,  the $\omega$,
after $\chi^2$-fitting, takes on the form
\begin{equation}
\omega = \frac{\sum_{ L'_{1} L'_{2} }(C_{ L'_{1} }+C_{ L'_{2}})
\jju }{2\sum_{ L'_{1} L'_{2}} \jju }
\end{equation}
where X is a 9j-symbol. Note that unlike $\alpha$ in (9), $\omega$ is not a
constant depends on $L_{1}L_{2}L$.  This means that for the various bands, it
will have different  contributions to the elementary excitations. In addition,
since it depends on L,  $\omega$ can cover more than the term $\alpha+\gamma(
\jjc -2j(j+1))$ in (8)  which is only limited to the rigid rotor case.

For any general interaction $C_{ \jj }$, the $\omega$ in (25) is usually
computed numerically. However, for QQ, we are able to find analytical
solutions for the two pair case.

\newcommand{\cat}{(2L_{1}+1)(2L_{2}+1)\sum_{k}(-1)^{k}(2k+1) }
\newcommand{\tea}{1+ \delta_{L_{1}L_{2}} - 2(2L_{1}+1)(2L_{2}+1)
X(jjL_{1};jjL_{2};L_{1}L_{2}L)}
\newcommand{\ka}{L_{1}}
\newcommand{\kb}{L_{2}}
\newcommand{\xx}{\frac{1}{2}}
\newcommand{\sat}{10(2L_{1}+1)(2L_{2}+1)F_{2}}
\newcommand{\lot}{ \cat C_{k}^{QQ}F_{k}}
\newcommand{\pot}{  W( \ka j \ka j;j2) W( \kb j \kb  j;j2) W(  \ka
\kb \ka \kb;L2)}
\newcommand{\mat}{10(2L_{1}+1)(2L_{2}+1) \pot }
\newcommand{\kat}{L_{1}(L_{1}+1)}
\newcommand{\kbt}{L_{2}(L_{2}+1)}
\newcommand{\kas}{L_{1}^{2}(L_{1}+1)^{2}}
\newcommand{\kbs}{L_{2}^{2}(L_{2}+1)^{2}}
\newcommand{\kit}{L_{i}(L_{i}+1)}
\newcommand{\kjt}{L_{j}(L_{j}+1)}
\newcommand{\kis}{L_{i}^{2}(L_{i}+1)^{2}}
\newcommand{\kjs}{L_{j}^{2}(L_{j}+1)^{2}}

Using  $C_{  \jj }^{QQ}=10W(jjjj;2 \jj )$ for QQ, one can derive the following
expressions:
\begin{equation}
\omega^{QQ} = -\frac{ \sat + \lot } { \tea }
\end{equation}
where $F_{k}$ is given by
\begin{eqnarray}
F_{k} &=& W( \ka j \ka j;jk) W( \kb j \kb j;jk) W( \ka \kb \ka  \kb ;Lk)
\\ \nonumber
      & &+ W( \ka j \kb j;jk)^2 W( \ka \kb \kb \ka ;Lk).
\end{eqnarray}

For the QQ-interaction, $\omega^{QQ}$ turns out to be simple in which the
largest spin pairs are preferred to be chosen as optimum pairs. One
can numerically show that both the second term in the numerator and
the denominator, and the second term of Eq.(27) are all shown to be
negligible. Thus,
\begin{eqnarray}
\omega^{QQ} &=& -
\frac{10}{1+\delta_{L_{1}L_{2}}}(2L_{1}+1)(2L_{2}+1)\\ \nonumber
& & \times \pot
\end{eqnarray}
This expression will reproduce well the exact values for  the low-lying bands.
The detail comparisons with the exact results are given in Fig. 1.

With this $\omega$, we can now cast the total energy into an analytical form:
\begin{equation}
E =E_{o}+ \frac{\hbar^{2}}{2\Im} L(L+1) - \chi L^{2}(L+1)^{2}
\end{equation}
where $E_{o}$ is the intrinsic energy given by
\newcommand{\pat}{15[ \kas +\kbs ]-15(4j^{2}+4j-1)[ \kat  + \kbt ]}
\newcommand{\sq}[2]{\bar{#1}^{#2}}
\newcommand{\qat}{3\sq{L_{1}}{2}-4\sq{j}{2}-3}
\newcommand{\rat}{3\sq{L_{2}}{2}-4\sq{j}{2}-3}
\newcommand{\xat}{ \jja  (2L_{1}-1)(2L_{2}-1)(2L_{1}+3)(2L_{2}+3)}
\newcommand{\dat}{3 \jjc -4j(j+1)-3}
\begin{equation}
E_{o} = C_{L_{1}}^{QQ}+C_{L_{2}}^{QQ}
-\chi \left[ \sq{L_{1}}{4} +  \sq{L_{2}}{4}  -\sq{L_{1}}{2}
-
\sq{L_{2}}{2}+ \frac{2}{3}\sq{L_{1}}{2}\sq{L_{2}}{2}\right]
\end{equation}
with the $\chi$ in the form
\begin{equation}
\chi = \frac{15[ \qat ][ \rat ]}{ \xat }
\end{equation}
and $\sq{L_{i}}{ }=\sqrt{L_{i}(L_{i}+1)}$,
$\sq{j}{ }=\sqrt{j(j+1)}$.
The moments of inertia are found to be
\begin{equation}
\frac{\hbar^{2}}{2\Im}  =  \chi [2 \sq{L_{1}}{2}
+2\sq{L_{2}}{2} -1]
\end{equation}
The  term $\chi L^{2}(L+1)^{2}$ which causes deviation from the rigid rotor
expression is in fact negligible for the QQ-interaction.

Since the choice of $\omega^{QQ}$ taken from a four-particle system is more
accurate
than those of $\alpha$ and $\gamma$ in (9), we now return to the rotational
limit of the many particle system (13)  and seek a better set of values them,
extracted from Eq.(29).  We thus set
\begin{equation}
C_{L_{1}}^{QQ}+C_{L_{2}}^{QQ}+4\alpha
- \gamma \left[ \kat + \kbt \right]=E_{o}
\end{equation}
from which one can obtain $\alpha$ and $\gamma$:
\[\alpha =\frac{\chi}{4}\left[\sq{L_{1}}{4}+\sq{L_{2}}{4}
+\frac{10}{3}\sq{L_{1}}{2}\sq{L_{2}}{2}\right]\]
\begin{equation}
\gamma =\frac{\hbar^{2}}{2\Im}
=  \chi \left[2 \sq{L_{1}}{2} +2\sq{L_{2}}{2} -1\right]
\end{equation}
Since these expressions are extracted from the nearly exact
solutions of the four particle problem, their ability to reproduce the inverted
SDI-like behavior, as given by Eq.(8), is a confirmation of the physics so far
discussed. Together with the former ones, the results of the moments of inertia
and the various band heads, calculated according to this choice of parameters,
are presented in Table I (listed as 2-Pair Fit). The values of $C'_{\Lambda}$
are listed in Table II. It is interesting to observe that these results are
indeed close to the inverted SDI ones and the exact SM. Of course, it remains
to be seen whether these expressions can be generalized to the system
of an arbitrary number of pairs.

In conclusion,  we have presented evidence in this paper that the excitations
induced by the QQ-interaction is equivalent to the inverted surface-delta
interaction  whose multipole-pairing component is dominant and attractive for
the largest multipole. This thus provide an understanding of the origin of the
recently introduced optimum pair description of rotation. Finally, it
should be emphasized that the present approach possesses a predictive power
for the excited bands using a simple Tamm-Dancoff approximation.

Research reported in this paper is supported by the Chinese National Natural
Science Foundation and the United States National Science Foundation.

\clearpage
\centering {\ Table I \\}
\ \\
\ \\
\centering {Inverted SDI vs EXACT SM}
\ \\
\vspace{7 mm}
\begin{tabular}{||c|c|c|c|c|c||}
\hline\hline
{        $Methods$         }  &  {\   $C'_{20}$}  &   {\
$\alpha$}  & {\  $\gamma$} & {\ $\gamma -E$} &   {\
$\beta -E$}  \\ \hline

{\       Approx.I} & {\     -1.95} & {\    0.84}
&  {\  $5.4 \times 10^{-3}$} & {\     0.68}  &
{\
1.36}  \\ \hline
{\    Approx.II} & {\    -2.08} & {\
0.91}
&  {\ $5.6 \times 10^{-3}$} & {\     0.70}  &
{\
1.40}  \\ \hline
{\   Inverted SDI} & {\     -1.53} & {\
0.62}
&  {\  $4.3 \times 10^{-3}$} & {\    0.59}  &
{\
1.18}  \\ \hline
{\     2-Pair Fit} & {\    -1.49} & {\
0.59}
&  {\  $4.2 \times 10^{-3}$} & {\     0.62}  &
{\
1.12}  \\ \hline
{\       Exact SM} & {\     -1.47} & {\
0.58}
& {\** $4.2 \times 10^{-3}$} & {\     0.62}  &
{\
1.12}  \\ \hline
\end{tabular}
\vspace{7 mm}

\newcommand{\hb}{\frac{\hbar^2}{2\Im}}
\centering ** Note: In Exact SM and the 2-Pair Fit, the moments of inertia
vary
 as follows:\\
\centering (a) $\hb $ =4.2 x $10^{-3}$ for g-band \\
\centering (b) $\hb $ =3.3 x $10^{-3}$ for $\gamma$-band \\
\centering (c) $\hb $ =2.5 x $10^{-3}$ for $\beta$-band

\clearpage
\centering {\ Table II \\}
\ \\
\ \\
\centering {\ The Values of $C'_{\Lambda}$}

\vspace{7 mm}

\begin{tabular}{||c|r|r|r|r||}                   \hline\hline
{\ $\Lambda$} & {\ I}  & {\ II}  & {\ III} & {\
IV} \\
 \hline
{\ 0} & {\ -0.001} & {\     -0} & {\ -0.042} & {\
 -0.029} \\ \hline
{\ 2} & {\      0} & {\ -4.E-4} & {\ -0.034} & {\
 -0.020} \\ \hline
{\ 4} & {\      0} & {\ -0.005} & {\ -0.019} & {\
 -0.004} \\ \hline
{\ 6} & {\ -0.009} & {\ -0.021} & {\ -0.019} & {\
 -0.014} \\ \hline
{\ 8} & {\ -0.040} & {\ -0.061} & {\ -0.004} & {\
 -0.018} \\ \hline
{\ 10} & {\ -0.110} & {\ -0.142} & {\ -0.003} & {\
 -0.006} \\ \hline
{\ 12} & {\ -0.239} & {\ -0.286} & {\ -0.110} & {\
 -0.081} \\ \hline
{\ 14} & {\ -0.456} & {\ -0.519} & {\ -0.267} & {\
 -0.232} \\ \hline
{\ 16} & {\ -0.786} & {\ -0.870} & {\ -0.532} & {\
 -0.490} \\ \hline
{\ 18} & {\ -1.273} & {\ -1.376} & {\ -0.939} & {\
 -0.891} \\ \hline
{\ 20} & {\ -1.948} & {\ -2.075} & {\ -1.529} & {\
 -1.473} \\ \hline\hline
\end{tabular}
\vspace{7 mm}
\ \\
\ \\
{\large
\centering Note. Column labels I,II,III and IV represent the following: \\
\centering I. Approximation I \\
\centering II. Approximation II \\
\centering III. Inverted SDI \\
\centering IV. Two Pair Fit}

\end{document}